\documentclass[smallextended]{svjour3}       
\smartqed  
\usepackage{graphicx,natbib,color,amsmath,amssymb,marvosym,fix-cm,mathptmx,bm}
\newcommand{\envelope}{(\raisebox{-.7pt}{\scalebox{1.45}{\Letter}}\kern0.1pt)}
\setcitestyle{aysep={}}
\usepackage[colorlinks=true,allcolors=blue]{hyperref}%

\begin{document}

\title{Education journal rankings: A diversity-based Author Affiliation Index assessment methodology}

\author{Yan-Hong Yang \and  Ying-Hui Shao}

\institute{Yan-Hong Yang \at SILC Business School, Shanghai University, Shanghai 201899, China
	\and
	Ying-Hui Shao \envelope \at School of Statistics and Information, Shanghai University of International Business and Economics, \\Shanghai 201620,China\\\email{yinghuishao@126.com}}

\date{Received: 18 August 2021/ Accepted:}
\maketitle
\begin{abstract}
  Determining the reputation of academic journals is an crucial issue. The Author Affiliation Index (AAI) was proposed as a novel indicator for judging journal quality in many academic disciplines. Nevertheless, the original AAI has several potential limitations, some of which have been discussed and addressed in previous studies. In this paper, we modified the original AAI by incorporating diversity of top-notch institutions, namely the AAID, exploring how institutional diversity is related to journal quality assessment. We further conducted a quality assessment of 263 education journals indexed in the Social Sciences Citation Index (SSCI) by applying the AAID, AAI and weighted AAI. We find that the AAID ranking possesses a low correlation coefficient with the Journal Impact Factor (JIF) and Eigenfactor Score (ES). That is to say, the AAID rating has not reached a good agreement with the most popular ranking indicators JIF and ES for journals in the field of education. Moreover, we analyze the reasons for the highest AAID from the structure of complex networks. Overall, the AAID is an alternative indicator for evaluating the prestige of journals from a new perspective.
\end{abstract}

\section{Introduction}
\label{Sec:Introduction}

Academic journal rankings have been disputable with the exception of a very small percentage of journals that can be unambiguously considered as top tier \citep{chen2007author}, nonetheless, they frequently serve as a broad proxy for research quality and its impact. Undoubtedly, journal rankings are also a vital benchmark regarding  promotion, hiring, remuneration  and  research  funding in academia \citep{currie2011finance,bajo2020should}.

A plethora of studies have been devoted to assessing journal quality. Methodologically, the two prevailing ways for journal rankings may be broadly categorized as citation-based methods and survey-based assessment \citep{garfield1955citation,hirsch2005index,currie2011finance,mingers2017evaluating, currie2020finance}. The JIF is a typical indicator evaluating the quality of academic journals based upon journal citations. The Journal Citation Reports (JCR) publish journal impact factor annually. For a long time, the JIF was the predominant journal metric despite considerable criticism. Later, there appeared several well-known indicators including the h-index, SJR, SNIP, ES and so forth, in which journal citation is the first and most basic measure. Because of different factors and particular biases, it is revealed that no one citation-based indicator is superior \citep{mingers2017evaluating}.

The survey-based assessment approach is another conventional way that journal ranking can be conducted by the perceived quality of a sample of experts in a particular discipline. Previous studies that have adopted this method include those by \cite{nederhof1991quality}, \cite{currie2011finance}, \cite{peters2014experts} and so on. Recently, \cite{currie2020finance} construct finance journal rankings and classifications using the active scholar assessment methodology with nested effects regression estimation. Nevertheless, the survey approach generally suffers from some limitations including questionnaire design biases, respondents' perceptions biases, and sampling biases \citep{moosa2011demise,chan2012accounting}.

In addition, scholars put forward other comprehensive methods to rank journal quality. For instance, \cite{bean2005estimating} proposed a parsimonious model that a journal's quality rating is a function of the age, acceptance rate, and classification of the journal as academic or professional. Whereafter, \cite{matherly2009pragmatic} improved the model proposed by \cite{bean2005estimating} via considering audience, journal availability, inclusion in the SSCI, and the journal's submission fee. Recently, \cite{bajo2020should} proposed a novel methodology to objective finance journal ranking by considering the impact of journal publications on career advancement. They present that the top three journals are significant drivers of promotion success, other journals are nearly as important, particularly for business schools outside of the top tier. Moreover, \cite{docampo2021journal} have proposed a new method, called paper afliation index,with which to rate journals based on bibliometric data.

The above-mentioned methodologies have their own merits as well as limitations and it is not the purpose of this study to delve into this topic. The intent of this paper is to concentrate on an author affiliation-based methodology that can be used to evaluate a wide variety of discipline specific journals. The Author Affiliation Index (AAI) is a novel indicator for assessing journal quality which was originally developed by \cite{harless1998revision}. Subsequent to \cite{harless1998revision}, \cite{gorman2005evaluating} expand the application of the AAI to the evaluation and ranking of a set of operations management journals. Since then, the AAI has been
broadly employed to evaluate diverse discipline specific journals \citep{chen2007author, ferratt2007journal, gorman2011evaluating, chan2012accounting, fry2014exploring, ginieis2020ranking}.

The philosophy of the AAI is as follows: to maintain or enhance their reputation capital, elite institutions prefer to employ productive faculty who achieve high-quality and influential researches \citep{ferratt2007journal}. Then, with the publication of more and more influential academic articles in certain journals, these journals become more prestigious in quality. In other words, a prestigious journal only publish high-quality research. Hence, the reputable journals have higher authorship concentration from top-ranking universities \citep{chen2007author}. Essentially, a journal's AAI can be constructed in view of the proportion of the journal's articles authored by scholars affiliated with elite programs \citep{chen2007author}. Therefore, the higher the AAI for a journal the higher is its corresponding rating.

Moreover, \cite{agrawal2011theoretical} present several concerns for employing the AAI as a viable alternative to surveys and citation-based approaches. To remedy potential problems expressed by \cite{agrawal2011theoretical}, \cite{fry2014exploring} then modified AAI by incorporating several improvements to the original AAI. The modified AAI mainly introduces a concept of the mean article AAI, a time dimension, and a weighting factor for an expanded set of elite institutions \citep{fry2014exploring}.

It has been reported that the institutional diversity play a critical role in the improvement of journal quality \citep{wu2020does}.
Even though growing attention has been paid to the AAI method, there remains some deficiency in calculating AAI. The original AAI method only focuses on the number of elite programs without taking into account their diversity \citep{gorman2005evaluating,chen2007author,fry2014exploring}. Therefore, the present paper builds a diversity-based Author Affiliation Index, and uses it as an alternative indicator to assess education journals' reputation.

The rest of this work is organized as follows. Section~\ref{Sec:Materials} describes the methodology and data. Section~\ref{Sec:Results} presents the empirical results. Conclusions and discussion are presented in the last section.

\section{Materials and Methods}
\label{Sec:Materials}
\vspace{6pt}

\subsection{The original AAI}

The original AAI formula was devised by \cite{harless1998revision} aiming to set a minimum standard for publication outlets for the faculty at Virginia
Commonwealth University, which was gradually regarded as an alternative measures of academic journal's reputation \citep{gorman2005evaluating,chen2007author, ferratt2007journal, gorman2011evaluating, chan2012accounting, fry2014exploring, ginieis2020ranking}. The formulation of AAI can be written as

\begin{equation}
AAI_k= \frac{\sum_{i\in M_k}X_i/N_i}{\sum_{i\in M_k}(X_i+Y_i)/N_i},
\label{Eq:AAI}
\end{equation}
where $AAI_k$ measures the Author Affiliation Index for journal $k$, and $AAI_k$ varies in the closed interval [0, 1]. $X_i$ is the number of authors from a set of elite universities in article $i$. $Y_i$ is the number of authors in article $i$ not from the top university set. $N_i$ is the total number of authors in article $i$, and $M_k$ represents the sample articles drawn from each journal $k$. Naturally,
the higher the AAI for a journal the higher is its corresponding ranking.

As argued in the previous studies, it is vital to determine the composition of the sample of top-notch universities \citep{chen2007author,fry2014exploring,ginieis2020ranking}. Following \cite{ginieis2020ranking}, we include academic authors from all top-tier universities globally, not just from the United States.

\subsection{Data Sets}
\label{Data}
\vspace{6pt}

To our knowledge, the AAI has not been yet used in the discipline of education and educational research. Hence, we apply the AAI method to assess the prestige of journals related to education and educational research. According to the 2019 InCites Journal Citation Reports, there are totally 263 journals in SSCI selected categories of `EDUCATION $\&$ EDUCATIONAL RESEARCH'. Then, the papers of the 263 education journals we analyze were drawn from the Web of Science (WoS) Core Collection database over the period from 1 January 1965 to 14 October 2020. In total, hundreds of thousands of papers are retrieved.
It is worth highlighting that only ``Article'', ``Proceedings Paper'',``Review'' and ``Book Chapter'' are taken into consideration in this study. Besides, the summary table of the 2019 JCR has reported JIF and ES for each journal, which are the benchmark for AAI and AAID.

When it comes to the selection of the set of top-notch institutions in the subject of education, we draw the institution ratings data from the QS Top University rankings (https://www.topuniversities.com, accessed on 1 Jun 2021). There exist four well-known ranking systems for the best global university, which are respectively the QS Top University rankings, the U.S. News $\&$ World Report Best Global Universities Rankings, the Times Higher Education World University Rankings, and the Academic Ranking of World Universities. With these four rating lists have their own ranking criteria, there is no clear-cut consensus on which ranking method is better. Following many academics, we here use the QS World University Rankings by subject of education, in which several criteria such as academic reputation, citations per faculty, student/faculty ratio, among others are taken into consideration \citep{fry2014exploring,ginieis2020ranking,hernandez2019technological,pilishvili2018top,calvo2017analysis}.

As reported in previous studies, $M_k$, the number of articles retrieved from each journal $k$, is set to equal 60 since the AAI becomes stabilized when the size of sample articles $M_k$ above 50 \citep{gorman2005evaluating,cronin2008applying,chen2007author,guthrie2012evaluating}. Actually, it's hard to draw 60 sample articles from each journal in only one year. Then, starting from the latest issue of 2020, we extract the sample articles in reverse chronological until 60 articles are picked.

\begin{figure}[!ht]
\centering
  \includegraphics[width=8.5cm]{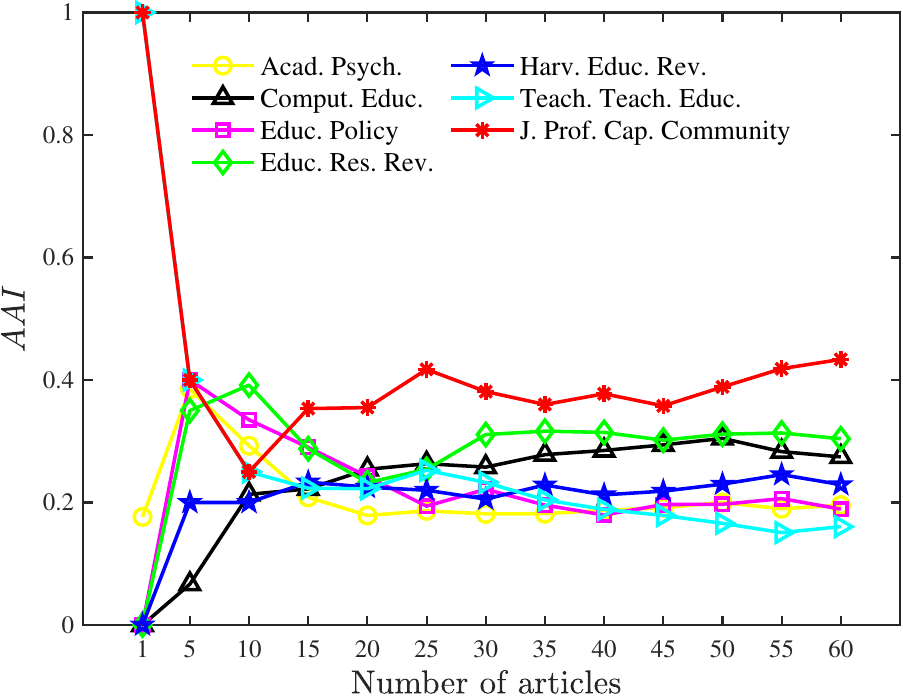}
  \caption{Illustration of AAI convergence behavior as size of sample articles increases. This figure depicts the AAIs for seven education journals which are respectively \emph{Academic Psychiatry} (Acad. Psych.), \emph{Computers $\&$ Education} (Comput. Educ.), \emph{Educational Policy} (Educ. Policy), \emph{Educational Research Review} (Educ. Res. Rev.), \emph{Harvard Educational Review} (Harv. Educ. Rev.), \emph{Teaching and Teacher Education} (Teach. Teach. Educ.) and \emph{Journal of Professional Capital and Community} (J. Prof. Cap. Community).}
  \label{Fig:AAI:Convergence}
\end{figure}

The convergence of AAI for education journals is demonstrated in Fig.~\ref{Fig:AAI:Convergence}, in which seven education journals were selected. One can observe that the AAIs of all these seven journals become considerably stable when $M$ is greater than 30, showing that our select of 60 sample articles to calculate AAI is reasonable. Specially, these seven journals contain both prestigious journal such as \emph{Harvard Educational Review} and new journal including \emph{Journal of Professional Capital and Community}.

Besides, we draw 60 sample articles from the majority of 263 education journals in two years. However, there still exist a large proportion of journals that need three years or above to reach 60 sample articles. In this case, we first retrieve the top 50 universities ranked by subject of education annually from 2017 to 2020. There exist a slight fluctuation of the top 50 universities every year.
Then, we filter out universities that appear only once in the top 50 universities from 2017 to 2019, and label the rest universities as set $B$. Hence, the union of the top 50 universities in 2020 (set $A$) and set $B$ is the final elite university sets, in which there will be a gathering of 58 top universities. We report the prestigious 58 universities in the education field in Table~\ref{Tb:TopUnivs:location}. Apparently, universities in the United States and the United Kingdom dominate the research of education.

\subsection{The weighted AAI}

In fact, for any preselected set of top-notch institutions, the contribution of the first ranked institution and the lowest ranked institution to the calculation of AAI is equal. \cite{moore1972relative} first proposed the idea of heavy weighting the most prestigious universities. Subsequently, \cite{fry2014exploring} introduced a weighting factor into the calculation of AAI, in which it gives more weight to authors from more highly regarded institutions than to authors from less highly
regarded institutions. The weighted AAI (AAIW) is depicted as follows

\begin{equation}
AAIW_k^{\rm weight}= \frac{\sum_{i\in M_k}w(ij) X_i/N_i}{\sum_{i\in M_k}(X_i+Y_i)/N_i},
\label{Eq:WeightedAAI}
\end{equation}
where $w(ij)$ is the weight coefficient of the affiliation of author $j$ of article $i$, and the rest parameters keep same as Eq. (\ref{Eq:AAI}). We refer to the AAIW mainly for the purpose of comparative analysis with the original AAI. We use the union of top 10 universities ranked by subject of education annually from 2017 to 2020 as the first tier institutions, which totally pick 13 the most prestigious universities. The 13 most prestigious universities are in bold in Table~\ref{Tb:TopUnivs:location}.
In this study, the 13 first tier institutions were weighted 1.2 times as heavily as the rest of all 58  top-notch institutions, which means weight coefficients $w(ij)$ of the 13 first tier institutions are set to 1.2. Naturally, the weight coefficients $w(ij)$ of the 45 second tier institutions is 1.

\subsection{The diversity-based AAI}

{\color{red}{Diversity is a crucial concept that is widely featured in numerous diverse disciplines, including economics \citep{nguyen2005variety}, ecology \citep{mccann2000diversity}, social sciences \citep{eagle2010network,jones2018change}, and of course scientometrics \citep{leydesdorff2019interdisciplinarity,rousseau2019leydesdorff,leydesdorff2018diversity,mutz2022diversity}. 
A general framework for defining diversity is proposed by \cite{stirling2007general}, where diversity is made up of variety, balance, and disparity. In other words, variety, balance, and disparity can be used to measure diversity, whether it is a combination of these three components or each one separately \citep{stirling2007general,leydesdorff2019interdisciplinarity,wu2020does}. However, the original AAI is a relatively simplified index that only considers variety and ignores balance. Meanwhile, balance is a very important factor when using AAI to measure the reputation of a journal because it can represent the  ``evenness'' of top institution considered \citep{stirling2007general,wu2020does}. Hence, we assume that, considering everything else being equal, the more even the balance, the more prestigious the journal will be. \citep{stirling2007general,wu2020does}.

As discussed above, diversity of the top universities is critical to measure the variety or balance of a fixed journal \citep{stirling2007general,wu2020does}. Thus, we modified the original AAI formula by multiplying the diversity $D_k$ of journal $k$ \citep{eagle2010network}, which can be quantified as a function of the Shannon entropy

\begin{equation}
  D_k=-\sum_{j=1}^{h}p_j \log(p_j),
  \label{Eq:Diversity}
\end{equation}
where $p_j$ is the proportion of authors affiliated with the top university $j$ in 60 sample articles, log $(p_j)$ is a natural logarithm, and $h$ is the size of the top universities.
Specially, a value of $D_k=0$ means that  one of the institutions has published all 60 sample
articles, i.e. one of the $p_j $ is equal to 1.
We then define a diversity-based Author Affiliation Index, $AAID_k$, as follows
\begin{equation}
  AAID_k=AAI_kD_k,
  \label{Eq:AAID}
\end{equation}
where higher $AAID_k$ indicates that the authors in journal $k$ affiliate with a variety of top-notch institutions. When AAI involved both weight and diversity (AAIWD), we rewrite Eq. (\ref{Eq:AAID}) as follows
\begin{equation}
  AAIWD_k=AAIW_kD_k.
  \label{Eq:AAIWD}
\end{equation}
 where $AAIWD_k$ is a comprehensive index and presented in Table~\ref{Tb:AAIW:Top13:45}.}}

\section{Results}
\label{Sec:Results}

\begin{table}[!htp]
\caption{
{Education journals ranked by the AAI and AAID. The top 58 universities reported in Table~\ref{Tb:TopUnivs:location} were involved in calculation of the AAI. $R$ is the corresponding ranking for each indicator.}}
\label{Tb:AAI:Top58}
\resizebox{\textwidth}{!}{
\centering
   \begin{tabular}{lccccccccccccccccccccc}
   \hline\hline
  Journal & AAI & $R^{\rm AAI}$ &  $D$ &  $R^{D}$  & AAID &  $R^{\rm AAID}$ & JIF& $R^{\rm JIF}$& ES& $R^{\rm ES}$\\
   \hline
 Journal of Professional Capital and Community & 0.434 &  1 & 2.906 &  3 & 1.260 &  1 & 0.824 & 218 & 0.00019 & 242\\
 British Educational Research Journal & 0.388 &  2 & 2.681 & 20 & 1.041 &  2 & 1.752 & 114 & 0.00220 & 50\\
 Discourse-Studies in the Cultural Politics of Education & 0.388 &  3 & 2.548 & 45 & 0.988 &  5 & 1.729 & 117 & 0.00250 & 43\\
 Australian Educational Researcher & 0.386 &  4 & 1.970 & 185 & 0.761 & 21 & 1.559 & 136 & 0.00078 & 164\\
 Learning Media and Technology & 0.360 &  5 & 2.761 & 11 & 0.995 &  4 & 2.547 & 44 & 0.00133 & 98\\
 British Journal of Sociology of Education & 0.356 &  6 & 2.600 & 34 & 0.925 &  8 & 1.782 & 112 & 0.00246 & 46\\
 Comparative Education & 0.354 &  7 & 2.296 & 102 & 0.813 & 15 & 2.204 & 69 & 0.00122 & 107\\
 Oxford Review of Education & 0.350 &  8 & 2.551 & 44 & 0.893 & 10 & 1.421 & 159 & 0.00171 & 79\\
 Journal of Philosophy of Education & 0.347 &  9 & 1.954 & 187 & 0.679 & 34 & 0.813 & 220 & 0.00062 & 185\\
 Comparative Education Review & 0.338 & 10 & 2.563 & 40 & 0.868 & 11 & 2.246 & 67 & 0.00126 & 103\\
 International Journal of Computer-Supported Collaborative Learning & 0.338 & 11 & 2.866 &  5 & 0.968 &  6 & 4.028 &  9 & 0.00074 & 166\\
 Review of Research in Education & 0.338 & 12 & 2.340 & 87 & 0.790 & 17 & 4.667 &  6 & 0.00137 & 94\\
 Asia Pacific Journal of Education & 0.333 & 13 & 1.948 & 188 & 0.649 & 39 & 0.733 & 228 & 0.00058 & 193\\
 Curriculum Inquiry & 0.333 & 13 & 2.223 & 127 & 0.741 & 25 & 1.111 & 190 & 0.00066 & 179\\
 Teachers and Teaching & 0.326 & 15 & 2.843 &  6 & 0.928 &  7 & 2.345 & 55 & 0.00214 & 53\\
 Professional Development in Education & 0.326 & 16 & 2.796 &  9 & 0.911 &  9 & 1.531 & 138 & 0.00086 & 153\\
 TESOL Quarterly & 0.323 & 17 & 3.103 &  1 & 1.002 &  3 & 2.071 & 84 & 0.00256 & 41\\
 Language Learning & 0.320 & 18 & 2.636 & 27 & 0.843 & 13 & 3.408 & 21 & 0.00379 & 17\\
 Educational Researcher & 0.317 & 19 & 2.324 & 93 & 0.736 & 26 & 3.483 & 20 & 0.00660 &  7\\
 Higher Education Research \& Development & 0.310 & 20 & 2.296 & 103 & 0.712 & 29 & 2.129 & 78 & 0.00297 & 29\\
 Asia-Pacific Education Researcher & 0.310 & 21 & 1.868 & 204 & 0.578 & 58 & 0.744 & 227 & 0.00099 & 131\\
 Journal of Research on Educational Effectiveness & 0.305 & 22 & 2.531 & 48 & 0.771 & 20 & 3.375 & 22 & 0.00288 & 30\\
 Compare-A Journal of Comparative and International Education & 0.304 & 23 & 2.837 &  7 & 0.863 & 12 & 1.607 & 127 & 0.00140 & 92\\
 Educational Research Review & 0.304 & 24 & 2.449 & 64 & 0.745 & 23 & 6.962 &  2 & 0.00299 & 28\\
 Teachers College Record & 0.297 & 25 & 2.211 & 131 & 0.657 & 37 & 0.970 & 205 & 0.00392 & 16\\
 Journal of Education for Teaching & 0.294 & 26 & 2.018 & 180 & 0.594 & 54 & 1.483 & 149 & 0.00114 & 114\\
 Cambridge Journal of Education & 0.294 & 27 & 2.639 & 26 & 0.776 & 19 & 1.421 & 159 & 0.00110 & 119\\
 British Journal of Educational Technology & 0.292 & 28 & 2.715 & 16 & 0.792 & 16 & 2.951 & 31 & 0.00341 & 23\\
 Journal of Literacy Research & 0.291 & 29 & 2.238 & 122 & 0.651 & 38 & 2.255 & 64 & 0.00089 & 150\\
 Pedagogische Studien & 0.290 & 30 & 1.402 & 230 & 0.406 & 114 & 0.245 & 261 & 0.00008 & 257\\
 Assessment \& Evaluation in Higher Education & 0.289 & 31 & 2.167 & 144 & 0.626 & 48 & 2.320 & 56 & 0.00213 & 54\\
 Educational Philosophy and Theory & 0.289 & 31 & 2.359 & 81 & 0.682 & 33 & 0.773 & 222 & 0.00173 & 78\\
 Language Teaching & 0.286 & 33 & 2.632 & 28 & 0.753 & 22 & 3.714 & 11 & 0.00157 & 85\\
 Asia Pacific Education Review & 0.285 & 34 & 2.345 & 85 & 0.668 & 35 & 0.761 & 224 & 0.00064 & 180\\
 International Journal of Science Education & 0.279 & 35 & 2.785 & 10 & 0.778 & 18 & 1.485 & 146 & 0.00416 & 15\\
 Journal of Education Policy & 0.275 & 36 & 2.692 & 18 & 0.741 & 24 & 3.048 & 29 & 0.00267 & 35\\
 Computers \& Education & 0.274 & 37 & 2.967 &  2 & 0.814 & 14 & 5.296 &  4 & 0.01337 &  1\\
 Reading Research Quarterly & 0.270 & 38 & 2.338 & 89 & 0.631 & 44 & 3.543 & 17 & 0.00179 & 76\\
 American Educational Research Journal & 0.270 & 39 & 2.565 & 39 & 0.692 & 32 & 5.013 &  5 & 0.00571 &  9\\
 Mathematical Thinking and Learning & 0.269 & 40 & 2.651 & 23 & 0.713 & 28 & 1.074 & 195 & 0.00033 & 223\\
 Education Finance and Policy & 0.267 & 41 & 2.373 & 76 & 0.633 & 41 & 2.395 & 51 & 0.00216 & 51\\
 Linguistics and Education & 0.264 & 42 & 2.626 & 30 & 0.693 & 31 & 1.289 & 176 & 0.00107 & 122\\
 English in Australia & 0.264 & 42 & 1.432 & 229 & 0.378 & 125 & 0.250 & 258 & 0.00005 & 260\\
 Journal of Educational Change & 0.261 & 44 & 2.664 & 22 & 0.695 & 30 & 1.791 & 110 & 0.00077 & 165\\
 Advances in Health Sciences Education & 0.259 & 45 & 2.337 & 90 & 0.606 & 52 & 2.480 & 47 & 0.00421 & 14\\
 Language Policy & 0.258 & 46 & 2.512 & 50 & 0.649 & 40 & 1.383 & 165 & 0.00063 & 183\\
 AERA Open & 0.255 & 47 & 2.383 & 75 & 0.608 & 50 & 1.892 & 99 & 0.00273 & 33\\
 Journal of Psychologists and Counsellors in Schools & 0.255 & 48 & 2.248 & 117 & 0.573 & 60 & 0.676 & 231 & 0.00012 & 253\\
 Educational Review & 0.253 & 49 & 2.398 & 71 & 0.606 & 51 & 2.042 & 87 & 0.00134 & 96\\
 Journal of Higher Education Policy and Management & 0.252 & 50 & 2.489 & 54 & 0.628 & 47 & 0.939 & 208 & 0.00092 & 144\\
   \hline\hline
   \end{tabular}}
\end{table}

\subsection{Education journal rankings using AAI }

{\color{red}{We calculated the AAI and AAID scores for all 263 education journals. As an example, Table~\ref{Tb:AAI:Top58} only displays 50 journals in descending order of the AAIs. The AAI scores of all 263 journals vary widely, ranging from 0 to 0.434, and they primarily cluster between 0.004 and 0.350 with a mean value of 0.170.}} As shown in Table~\ref{Tb:AAI:Top58}, \emph{Journal of Professional Capital and Community} (hereafter referred to as \emph{JPCC}) ranks first with the highest score of 0.434, which means that 43.4\% of this journal's articles are authored by scholars from the top 58 ranking institutions in Table~\ref{Tb:TopUnivs:location}. In addition, \emph{JPCC} is the only journal with an AAI score above 0.4. Likewise, \emph{JPCC} has the highest AAID score of 1.260 and ranks third according to the diversity ratings. However, \emph{JPCC} is a new journal with the first issue published in 2016, thus its JIF and ES hold a low rating of 218
and 242 respectively. In contrast, while \emph{Review of Educational Research} ranks number one among all the 263 education journals with the highest JIF of 8.327, its ratings are 229, 222 and 206 ranked by the AAI, AAID and diversity respectively. Correspondingly, \emph{Computers $\&$ Education} possesses the highest ES, and ranks thirty-seventh, fourteenth, second using the AAI, AAID and diversity respectively. As reported in Table~\ref{Tb:AAI:Top58}, there exist inconsistencies in the rankings of education journals by using these indicators of AAI, AAID, JIF, ES and diversity.

In addition, a comparative analysis was conducted by shrinking the size of the set of top-notch institutions. In total, the top 13 universities were involved in calculation of the AAI and AAID. And Table~\ref{Tb:AAI:Top13} reports the top 50 journals ranked by the AAI. In Table~\ref{Tb:AAI:Top13}, \emph{Journal of Education for Teaching} holds the highest AAI at 0.225 while \emph{JPCC} ranks ninth. Moreover, \emph{JPCC} ranks second according to the AAID. Overall, it was noticeable that Table~\ref{Tb:AAI:Top13} shows a lower AAIs comparing with Table~\ref{Tb:AAI:Top58}.

When it comes to the weighted AAI, Table~\ref{Tb:AAIW:Top13:45} reports a detailed ranking of the top 50 education journals. Roughly, the AAIW in Table~\ref{Tb:AAIW:Top13:45} becomes higher comparing with the original AAI in Table~\ref{Tb:AAI:Top58}. And there are some minor inconsistencies in the ranking of education journals between Table~\ref{Tb:AAI:Top58} and Table~\ref{Tb:AAIW:Top13:45}. For instance, \emph{Learning Media and Technology} ranks fifth in Table~\ref{Tb:AAI:Top58} and ninth in Table~\ref{Tb:AAIW:Top13:45}, which indicates that the proportion of authors from extremely top institutions in \emph{Learning Media and Technology} is low.

\subsection{Comparisons: AAID with D, AAI, JIF and ES}

In this section, a quantitative comparison of the AAID journal rankings with other ranking systems such as the journal impact factor, eigenfactor score, AAI and diversity is implemented in Table~\ref{Tb:Comparisons}. Specifically, Table~\ref{Tb:Comparisons} gives the Spearman correlation coefficients among the five journal ranking indicators in three panels. And Panel A, Panel B and Panel C show correlation analysis results which correspond to values of indicators in Table~\ref{Tb:AAI:Top58}, Table~\ref{Tb:AAI:Top13} and Table~\ref{Tb:AAIW:Top13:45}, respectively.

As shown in Panel A of Table~\ref{Tb:Comparisons}, there exists a significant positive correlation between the AAID and other four indicators. Besides, it can be seen that the AAID ranking is highly correlated with the AAI and diversity, while the AAID possesses a low correlation coefficient of 0.269 with JIF, and 0.398 with ES.
Regarding the results in Panel B of Table~\ref{Tb:Comparisons}, it shows the correlation between the AAID and JIF is not statistically significant. However, one can observe a consistent result between Panel A and Panel B that the AAID presents a positive correlation coefficient of 0.278 with ES in Panel B. In addition, Panel C reports a similarly positive correlation between AAIWD and other indicators comparing with Panel A. Roughly speaking, the AAID ranking has not reached a good agreement with conventional journal ranking indicators JIF and ES.

\begin{table}[h]
\caption{Means, standard deviations, and Spearman correlation coefficients of five indicators. The superscripts*, ** and *** represents statistical significance at 0.05, 0.001 and 0.0001 levels, respectively.}
\label{Tb:Comparisons}
\centering
   \begin{tabular}{llcccccccccccccccccccc}
   \hline
 Order & Indicators & Mean& Std.Dev & 1 & 2 & 3 & 4 &  5 \\
   \hline
 \multicolumn{4}{l}{\textit{Panel A : Five indicators in Table~\ref{Tb:AAI:Top58}}} \\

1&JIF& 1.779 & 1.111 & 1\\
2&ES& 0.002 &0.002& 0.646*** & 1\\
3&AAI& 0.170 & 0.090 & 0.210**~~ &0.343***&1\\
4&D& 2.086 & 0.572 & 0.316***&0.429***&0.696***&1\\
5&AAID& 0.387 & 0.245 & 0.269**&0.398***&0.969***&0.826***&1\\
\hline
\multicolumn{4}{l}{\textit{Panel B : Five indicators in Table~\ref{Tb:AAI:Top13}}} \\

1&JIF& 1.779 & 1.111 & 1\\
2&ES& 0.002 &0.002& 0.632*** & 1\\
3&AAI& 0.058 & 0.045 & 0.094~~~~ &0.258***&1\\
4&D& 0.992 & 0.507 & 0.135*~~~&0.254***&0.662***&1\\
5&AAID& 0.070 & 0.071 & 0.101~~~~~&0.278***&0.934***&0.860***&1\\
\hline
\multicolumn{4}{l}{\textit{Panel C : Five indicators in Table~\ref{Tb:AAIW:Top13:45}}} \\

1&JIF& 1.779 & 1.111 & 1\\
2&ES& 0.002 &0.002& 0.646*** & 1\\
3&AAIW& 0.181 & 0.097 & 0.205**~~ &0.343***&1\\
4&D& 2.086 & 0.572 & 0.316***&0.429***&0.698***&1\\
5&AAIWD& 0.412 & 0.263 & 0.263***&0.396***&0.972***&0.824***&1\\
   \hline
   \end{tabular}
\end{table}

\subsection{Bibliometric analysis of \emph{JPCC}}

\begin{figure}[h]
\centering
  \includegraphics[width=11cm]{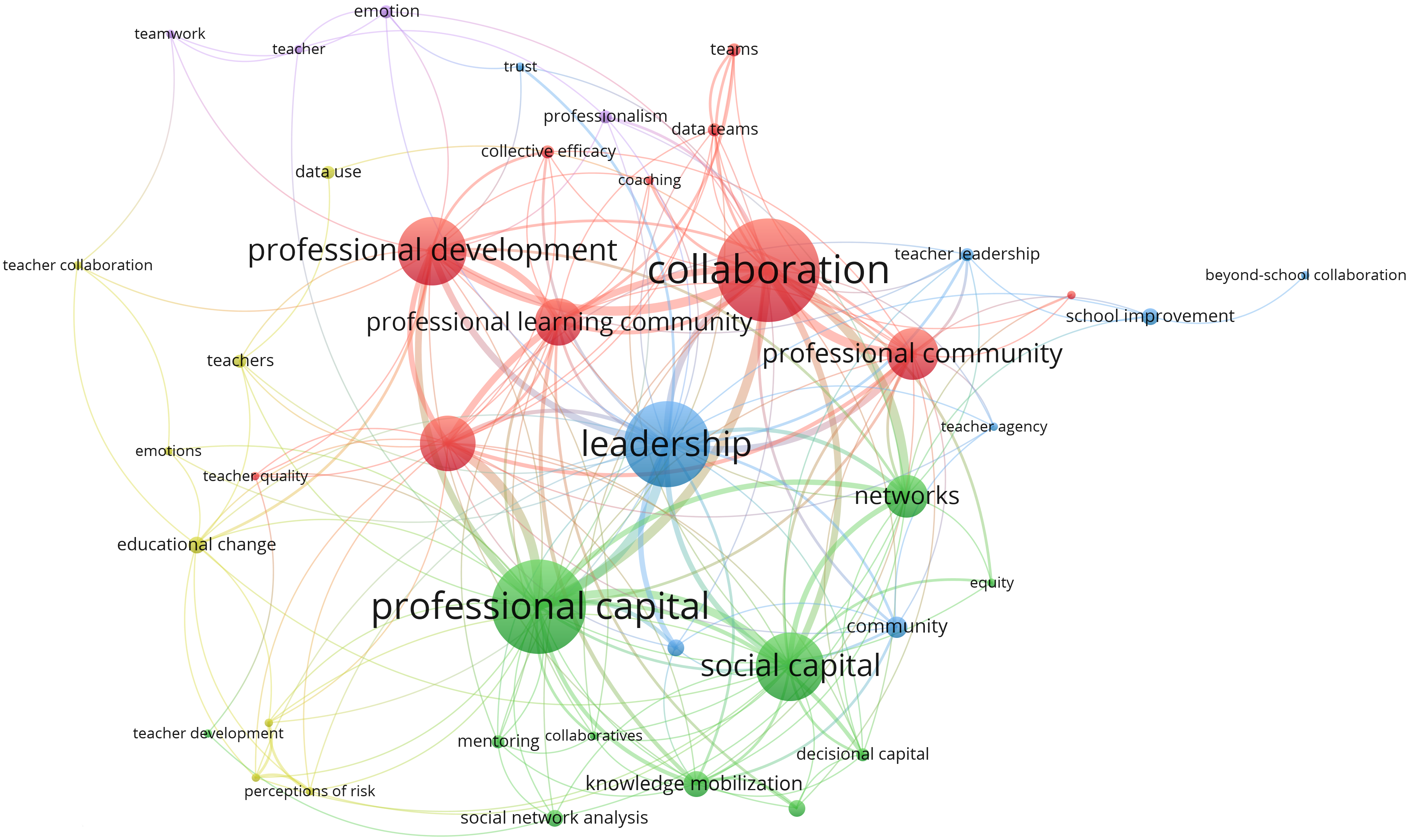}
  \caption{ Co-occurrence map of Author keywords (N = 208 keywords; threshold 2 co-occurrences,  display 42 keywords).}
  \label{Fig:JPCC:Co_occurrence}
\end{figure}

For an intuitive understanding of the highest place of \emph{JPCC} in the AAID and AAIWD rankings, a brief bibliometric analysis on \emph{JPCC} was conducted in this section. As descripted in section~\ref{Data}, a total of 81 publications from \emph{JPCC} were retrieved. Besides, there exist 211 citing papers for the 81 articles. For revealing topical themes of \emph{JPCC}, we first investigate the co-occurrence map of author keywords of the 81 papers. Then, using a bibliometric software (VOSviewer), we set a threshold of at least 2 co-occurrences and displayed 42 keywords on the co-occurrence network (see Fig. \ref{Fig:JPCC:Co_occurrence}). It is apparent that the most frequently co-occurring keywords in the 81 articles are Collaboration, Professional Capital, Leadership, Professional Development, Social Capital, Professional Community and themes relating to teaching. That is to say, \emph{JPCC} is
an extremely specialized journal focusing on themes of professional capital and education profession. Therefore, \emph{JPCC} will stand a good chance of attracting top scholars in these fields.

\begin{table}[!htp]
\caption{Institutions with the greatest output of papers in 81 publications from \emph{JPCC}. Note: $R$ denotes rank. TP represents the number of publications affiliated a specific institution.  $\rho$ is the percentage of papers published by a specific institution. TC is the total citations of TP. PY is the average publication year of TP.}
\label{Tb:Corepaper:Institution}
\centering
   \begin{tabular}{llcccccccccccccccccccc}
   \hline\hline
  $R$ & Institution & TP &  $\rho$ &  TC  & TC/TP & PY  \\
   \hline
  1  & Univ Calif San Diego &  6 & 7.41\% & 41 &  6.83 & 2017.5 \\
  2  & Boston Coll &  6 & 7.41\% & 17 &  2.83 & 2017.3 \\
  3  & Univ Toronto &  4 & 4.94\% & 51 & 12.75 & 2017.2 \\
  4  & Univ Twente &  4 & 4.94\% & 14 &  3.50 & 2018.2 \\
  5  & UCL &  3 & 3.70\% & 35 & 11.67 & 2017.3 \\
  6  & Natl Taipei Univ Educ &  2 & 2.47\% & 25 & 12.50 & 2016.5 \\
  7  & Stanford Univ &  2 & 2.47\% & 23 & 11.50 & 2016.0 \\
  8  & Open Univ Cyprus &  2 & 2.47\% & 18 &  9.00 & 2017.0 \\
  9  & Vrije Univ Amsterdam &  2 & 2.47\% & 14 &  7.00 & 2017.5 \\
 10  & Univ Worcester &  2 & 2.47\% & 14 &  7.00 & 2018.0 \\
 11  & Univ Auckland &  2 & 2.47\% &  4 &  2.00 & 2018.0 \\
 12  & Mid Sweden Univ &  2 & 2.47\% &  1 &  0.50 & 2019.5 \\
 13  & Columbia Univ &  2 & 2.47\% &  1 &  0.50 & 2019.0 \\
 14  & Univ Antwerp &  2 & 2.47\% &  1 &  0.50 & 2018.5 \\
 15  & York Univ &  2 & 2.47\% &  1 &  0.50 & 2018.0 \\
 16  & Univ Manchester &  1 & 1.23\% & 23 & 23.00 & 2016.0 \\
 17  & Univ Glasgow &  1 & 1.23\% & 17 & 17.00 & 2016.0 \\
 18  & Univ Stirling &  1 & 1.23\% & 16 & 16.00 & 2016.0 \\
 19  & Univ Oxford &  1 & 1.23\% & 14 & 14.00 & 2016.0 \\
 20  & Michigan State Univ &  1 & 1.23\% & 13 & 13.00 & 2018.0 \\
   \hline\hline
   \end{tabular}
\end{table}

Besides, Table~\ref{Tb:Corepaper:Institution} presents institutions with the greatest output of papers in 81 publications from \emph{JPCC}. University of California San Diego, Boston College, University of Toronto, University of Twente and University College London are the top five institutions in terms of publications, accounting for 7.41\%, 7.41\%, 4.94\%, 4.94\% and 3.70\% of the 81 papers of \emph{JPCC}, respectively. In total, USA, UK, and Canada are the   the most productive countries publishing 41.98\%, 18.52\%, and 12.35\% of the 81 articles, respectively. According to Table ~\ref{Tb:Citingpaper:Institution}, the institution that accounts for the highest output of the citing papers is the University College London. And countries with the greatest output of citing papers are  USA, UK and Spain with percentages of 17.69\%, 17.33\% and 7.22\%, respectively. In other words, both publications and citing papers of \emph{JPCC} are favored by top institutional scholars in the field of education.

\begin{figure}[h]
\centering
  \includegraphics[width=8.5cm]{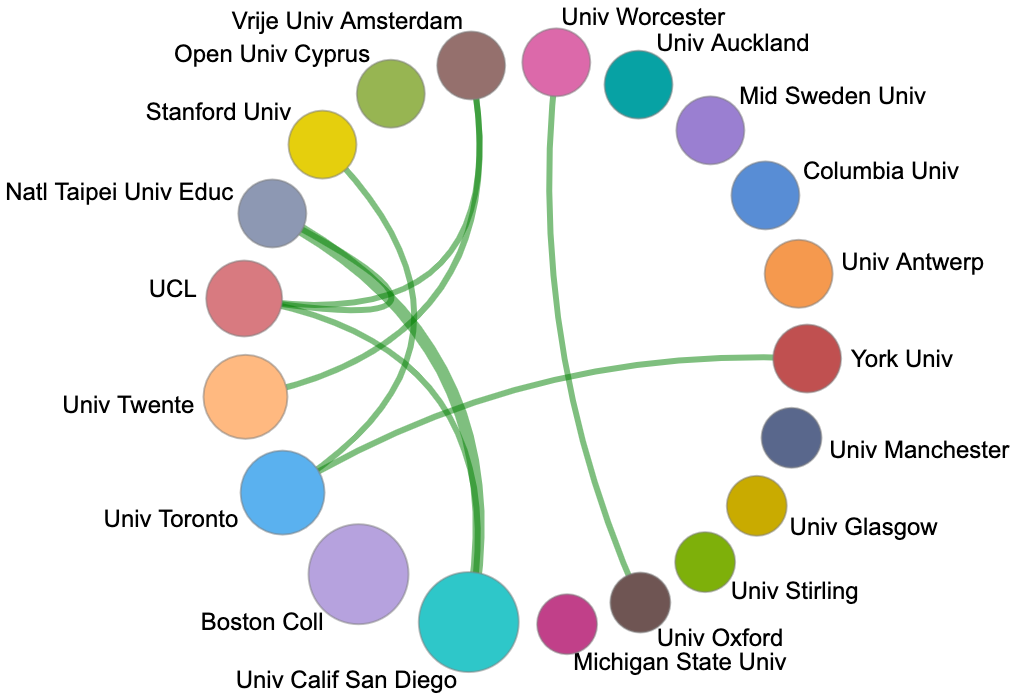}
  \caption{Collaboration network among major institutions of 81 papers from \emph{JPCC}.}
  \label{Fig:JPCC:Colla:Institution}
\end{figure}

The collaboration network among major institutions is shown in Fig. \ref{Fig:JPCC:Colla:Institution}. Sparse connection of collaboration network indicates that
the publications of \emph{JPCC} are usually collaborated among much smaller numbers of researchers. Actually, the vast majority of publications from \emph{JPCC} have a fewer number of co-authors, which can partly account for the higher AAID of \emph{JPCC} \citep{ginieis2020ranking}.

\section{Conclusions and Discussion}
\label{Sec:Conclusion}

It is certain that identifying the most prestigious journals in any academic field is an crucial issue and also a challenge. The introduction of AAI has created a new evaluation framework for academic journals. In this work, we improved the original AAI method and introduced a more comprehensive AAID indicator, exploring how institutional diversity is related to journal quality assessment. Based on these methods, we conducted an quality assessment of 263 education journals. To the best of our knowledge, this is the first study by applying the methods of AAI families to SSCI journals in the education field. In addition, comparing with the relatively few sample journals in other literatures, a larger set of 263 education journals were included for this study.

Our results indicate that the AAI scores of education journals fluctuate more greatly than that in the field of sustainability \citep{ginieis2020ranking}. Compared with other disciplines such as finance, accounting, transportation \citep{chen2007author,gorman2005evaluating,guthrie2012evaluating}, the high quantiles of AAI scores of education journals are relatively smaller. The co-author patters and other factors in different disciplines probably account for the different AAI values.

Besides, \emph{JPCC} has always been among the best in the ranking of various indicators of AAI families (including the original AAI, AAID, AAIW and AAIWD). We carry out a brief bibliometric analysis  to intutitively understand reasons for \emph{JPCC}'s high ranking. We found that the focusing on specific research topics and sparse collaboration patterns may contribute to \emph{JPCC}'s high rating.

We additionally conduct a quantitative comparison of the AAID rankings with other indicators reported in Table~\ref{Tb:Comparisons}. According to the comparative correlation analysis of these indicators, there are significant, positive correlations among the five ranking indices.
Nevertheless, it can be seen that the AAID ranking is highly correlated with the AAI and diversity, while the AAID presents a low correlation coefficient of 0.269 with JIF, and 0.398 with ES. In other words, the AAID ranking has not achieved a good consistency with the most popular ranking indicators JIF and ES for journals in the field of education.

The AAID provides an objective evaluation method with simple implementation, less time-consuming and easy to quantify. Therefore, the AAID can serve as an alternative ranking for journals besides JIF, ES and other measures. Meanwhile, the original AAI method possesses certain limitations which have been discussed and improved in previous studies \citep{fry2014exploring,ginieis2020ranking}. When it comes to the limitation of AAID method, the future improvements can be modified from the citing patterns, choosing the composition of and cardinality of top institutions and other aspects. In particular, the AAID can be applied to other fields such as finance and accounting to test the ranking performance in future research.

\begin{acknowledgements}
       This work was supported by the National Natural Science Foundation of China (12005064).  
\end{acknowledgements}

\section{Appendix}
\label{Sec:Appendix}

\begin{table}[ht]
\caption{
{ The top 58 universities worldwide in the disciplines of education and educational research. All bottom eight universities are the annually top 50 QS universities ranked by subject of education from 2017 to 2019, and tied for fifty-first place. The 13 most prestigious universities are in bold.}}
\label{Tb:TopUnivs:location}
\resizebox{\textwidth}{!}{
\centering
   \begin{tabular}{cllcll}
   \hline\hline
  Rank~2020 &University & Location &Rank~2020&University & Location \\
   \hline
  1  & \bf{University College London} & UK& 30 & Utrecht University & Netherlands \\
  2  & \bf{Harvard University} & USA& 31 & National Taiwan Normal University & Taiwan \\
  2  & \bf{Stanford University} & USA& 32 & Deakin University & Australia \\
  4  & \bf{University of Oxford} & UK& 33 & University of Pennsylvania & USA \\
  5  & \bf{University of Toronto} & Canada& 34 & The University of Manchester & UK \\
  6  & \bf{University of Cambridge} & UK& 35 & University of Illinois at Urbana-Champaign & USA \\
  7  & \bf{The University of Hong Kong} & China& 36 & University of Bristol & UK \\
  8  & \bf{University of California, Berkeley} & UK& 37 & Pontificia Universidad Catolica de Chile & Chile \\
  9  & \bf{University of  British Columbia} & Canada& 37 & University of Washington & USA \\
 10  & \bf{Columbia University} & USA& 39 & University of Texas at Austin & USA \\
 11  & \bf{University of California, Los Angeles} & USA& 40 & KU Leuven & Belgium \\
 12  & Monash University & Australia& 41 & University of Glasgow & UK \\
 12  & \bf{The University of Melbourne} & Australia& 42 & University of Chicago  & USA \\
 14  & University of Wisconsin-Madison & USA& 43 & The Chinese University of Hong Kong  & China \\
 15  & University of Michigan-Ann Arbor & USA& 44 & University of Alberta & Canada \\
 16  & Nanyang Technological University & Singapore& 44 & University of Birmingham & UK \\
 16  & \bf{The Education University of Hong Kong} & China& 46 & University of Oslo & Norway \\
 18  & Michigan State University & USA& 47 & New York University & USA \\
 19  & Johns Hopkins University  & USA& 48 & Seoul National University & South Korea \\
 20  & The University of Sydney & Australia& 48 & University of Leeds & UK \\
 21  & The University of Edinburgh & UK& 50 & Universidad Nacional Autonoma de Mexico & Mexico \\
 22  & Vanderbilt University & USA& 51 & University of Illinois at Chicago & USA \\
 23  & The University of Queensland & Australia& 51 & University of Amsterdam & Netherlands \\
 23  & University of Helsinki & Finland& 51 & University of  Nottingham & UK \\
 25  & Beijing Normal University  & China& 51 & Boston College & USA \\
 26  & McGill University  & Canada& 51 & University of Twente & Netherlands \\
 27  & The University of Auckland & New Zealand& 51 & The University of New South Wales & Australia \\
 28  & King's College London & UK& 51 & The Australian National University  & Australia \\
 29  & Pennsylvania State University & USA& 51 & Universidad de Chile & Chile \\
   \hline\hline
   \end{tabular}}
\end{table}
\newpage

\begin{table}[ht]
\caption{
{Education journals rankings using the AAI and AAID. The top 13 universities were involved in calculation of the AAID, and depicted in bold in Table~\ref{Tb:TopUnivs:location}. $R$ is the corresponding rating for each indicator.}}
\label{Tb:AAI:Top13}
\resizebox{\textwidth}{!}{
\centering
   \begin{tabular}{lccccccccccccccccccccc}
   \hline\hline
  Journal & AAI & $R^{\rm AAI}$ &  $D$ &  $R^{D}$  & AAID &  $R^{\rm AAID}$ & JIF& $R^{\rm JIF}$& ES& $R^{\rm ES}$\\
   \hline
 Journal of Education for Teaching & 0.225 &  1 & 1.271 & 86 & 0.286 &  5 & 1.483 & 149 & 0.00114 & 114\\
 Journal of Philosophy of Education & 0.222 &  2 & 1.230 & 95 & 0.273 &  6 & 0.813 & 220 & 0.00062 & 185\\
 Comparative Education & 0.207 &  3 & 1.268 & 87 & 0.262 &  7 & 2.204 & 69 & 0.00122 & 107\\
 Curriculum Inquiry & 0.194 &  4 & 1.488 & 39 & 0.289 &  4 & 1.111 & 190 & 0.00066 & 179\\
 British Educational Research Journal & 0.181 &  5 & 1.330 & 70 & 0.240 & 10 & 1.752 & 114 & 0.00220 & 50\\
 British Journal of Sociology of Education & 0.177 &  6 & 1.378 & 59 & 0.244 &  9 & 1.782 & 112 & 0.00246 & 46\\
 Comparative Education Review & 0.175 &  7 & 1.717 &  8 & 0.300 &  3 & 2.246 & 67 & 0.00126 & 103\\
 International Journal of Educational Development & 0.172 &  8 & 1.934 &  4 & 0.333 &  1 & 1.360 & 168 & 0.00237 & 48\\
 Journal of Professional Capital and Community & 0.168 &  9 & 1.986 &  2 & 0.333 &  2 & 0.824 & 218 & 0.00019 & 242\\
 Language Teaching & 0.154 & 10 & 1.642 & 14 & 0.253 &  8 & 3.714 & 11 & 0.00157 & 85\\
 Review of Research in Education & 0.151 & 11 & 1.127 & 105 & 0.170 & 24 & 4.667 &  6 & 0.00137 & 94\\
 Journal of Higher Education Policy and Management & 0.151 & 11 & 1.439 & 48 & 0.217 & 14 & 0.939 & 208 & 0.00092 & 144\\
 Oxford Review of Education & 0.150 & 13 & 1.273 & 84 & 0.191 & 20 & 1.421 & 159 & 0.00171 & 79\\
 Educational Philosophy and Theory & 0.142 & 14 & 1.089 & 117 & 0.154 & 30 & 0.773 & 222 & 0.00173 & 78\\
 Asia-Pacific Education Researcher & 0.142 & 14 & 0.637 & 188 & 0.090 & 82 & 0.744 & 227 & 0.00099 & 131\\
 Compare-A Journal of Comparative and International Education & 0.140 & 16 & 1.673 & 12 & 0.234 & 12 & 1.607 & 127 & 0.00140 & 92\\
 Teachers College Record & 0.138 & 17 & 1.194 & 101 & 0.165 & 26 & 0.970 & 205 & 0.00392 & 16\\
 Teachers and Teaching & 0.133 & 18 & 1.750 &  7 & 0.233 & 13 & 2.345 & 55 & 0.00214 & 53\\
 Asia Pacific Journal of Education & 0.132 & 19 & 1.091 & 116 & 0.144 & 35 & 0.733 & 228 & 0.00058 & 193\\
 Studies in Philosophy and Education & 0.125 & 20 & 1.386 & 52 & 0.173 & 23 & 0.650 & 232 & 0.00054 & 202\\
 Discourse-Studies in the Cultural Politics of Education & 0.123 & 21 & 1.696 & 10 & 0.208 & 17 & 1.729 & 117 & 0.00250 & 43\\
 Linguistics and Education & 0.122 & 22 & 1.609 & 18 & 0.197 & 19 & 1.289 & 176 & 0.00107 & 122\\
 Learning and Instruction & 0.122 & 23 & 1.772 &  6 & 0.216 & 15 & 3.323 & 24 & 0.00748 &  4\\
 TESOL Quarterly & 0.121 & 24 & 1.946 &  3 & 0.236 & 11 & 2.071 & 84 & 0.00256 & 41\\
 Sex Education-Sexuality Society and Learning & 0.119 & 25 & 1.523 & 34 & 0.182 & 22 & 1.480 & 152 & 0.00124 & 104\\
 Advances in Health Sciences Education & 0.117 & 26 & 0.872 & 155 & 0.102 & 65 & 2.480 & 47 & 0.00421 & 14\\
 AERA Open & 0.116 & 27 & 1.244 & 90 & 0.144 & 34 & 1.892 & 99 & 0.00273 & 33\\
 Scientific Studies of Reading & 0.115 & 28 & 1.712 &  9 & 0.197 & 18 & 2.910 & 32 & 0.00275 & 32\\
 Professional Development in Education & 0.112 & 29 & 1.668 & 13 & 0.187 & 21 & 1.531 & 138 & 0.00086 & 153\\
 Journal of the Learning Sciences & 0.112 & 30 & 1.517 & 35 & 0.169 & 25 & 3.588 & 16 & 0.00159 & 83\\
 ELT Journal & 0.108 & 31 & 1.474 & 43 & 0.160 & 27 & 1.314 & 173 & 0.00095 & 139\\
Sociology of Education & 0.106 & 32 & 1.418 & 50 & 0.150 & 31 & 3.647 & 14 & 0.00206 & 56\\
 International Journal of Science Education & 0.106 & 32 & 2.025 &  1 & 0.214 & 16 & 1.485 & 146 & 0.00416 & 15\\
 Journal of Social Work Education & 0.106 & 32 & 1.359 & 61 & 0.143 & 36 & 0.845 & 215 & 0.00079 & 163\\
 Assessment \& Evaluation in Higher Education & 0.101 & 35 & 1.280 & 78 & 0.129 & 48 & 2.320 & 56 & 0.00213 & 54\\
 Language Teaching Research & 0.100 & 36 & 1.468 & 45 & 0.147 & 32 & 2.647 & 40 & 0.00205 & 58\\
 Reading and Writing & 0.100 & 36 & 1.240 & 94 & 0.124 & 52 & 1.445 & 155 & 0.00379 & 17\\
 Race Ethnicity and Education & 0.100 & 36 & 1.561 & 22 & 0.156 & 29 & 1.807 & 108 & 0.00238 & 47\\
 Cambridge Journal of Education & 0.099 & 39 & 1.215 & 97 & 0.120 & 56 & 1.421 & 159 & 0.00110 & 119\\
 Language Learning & 0.097 & 40 & 1.386 & 52 & 0.135 & 46 & 3.408 & 21 & 0.00379 & 17\\
 Language Policy & 0.097 & 40 & 1.280 & 78 & 0.124 & 51 & 1.383 & 165 & 0.00063 & 183\\
 Higher Education Policy & 0.097 & 42 & 1.609 & 18 & 0.156 & 28 & 1.597 & 130 & 0.00060 & 190\\
 Academic Psychiatry & 0.097 & 43 & 1.205 & 100 & 0.117 & 57 & 2.148 & 75 & 0.00351 & 22\\
 Harvard Educational Review & 0.096 & 44 & 1.386 & 52 & 0.133 & 47 & 1.818 & 105 & 0.00325 & 26\\
 Higher Education & 0.095 & 45 & 1.265 & 88 & 0.120 & 55 & 2.856 & 33 & 0.00557 & 11\\
 English in Australia & 0.094 & 46 & 0.305 & 211 & 0.029 & 170 & 0.250 & 258 & 0.00005 & 260\\
 Education Finance and Policy & 0.094 & 47 & 1.030 & 135 & 0.097 & 72 & 2.395 & 51 & 0.00216 & 51\\
 International Journal of Art \& Design Education & 0.094 & 47 & 1.321 & 73 & 0.125 & 50 & 0.475 & 244 & 0.00009 & 255\\
 British Journal of Educational Studies & 0.094 & 49 & 1.003 & 139 & 0.094 & 77 & 1.604 & 128 & 0.00093 & 143\\
 Mathematical Thinking and Learning & 0.092 & 50 & 1.242 & 91 & 0.115 & 59 & 1.074 & 195 & 0.00033 & 223\\
    \hline\hline
   \end{tabular}}
\end{table}

\newpage

\begin{table}[ht]

\caption{Education journals rankings using the AAIW and AAIWD, where the top 13 ranked universities were weighted 1.2 times as heavily as the remainder of all 58 top-notch institutions. All the 13 highest ranked institutions and 45 less highly regarded institutions are reported in Table~\ref{Tb:TopUnivs:location} in which the 13 more highly regarded institutions are depicted in bold. $R$ is the corresponding ranking for each indicator.}
\label{Tb:AAIW:Top13:45}
\resizebox{\textwidth}{!}{
\centering
   \begin{tabular}{lccccccccccccccccccccc}
   \hline\hline
  Journal & AAIW & $R^{\rm AAIW}$ &  $D$ &  $R^{D}$  & AAIWD &  $R^{\rm AAIWD}$ & JIF& $R^{\rm JIF}$& ES& $R^{\rm ES}$\\
   \hline
 Journal of Professional Capital and Community & 0.464 &  1 & 2.906 &  3 & 1.348 &  1 & 0.824 & 218 & 0.00019 & 242\\
 British Educational Research Journal& 0.424 &  2 & 2.681 & 20 & 1.138 &  2 & 1.752 & 114 & 0.00220 & 50\\
 Discourse-Studies in the Cultural Politics of Education & 0.406 &  3 & 2.548 & 45 & 1.034 &  4 & 1.729 & 117 & 0.00250 & 43\\
 Australian Educational Researcher & 0.403 &  4 & 1.970 & 185 & 0.793 & 23 & 1.559 & 136 & 0.00078 & 164\\
 Comparative Education & 0.395 &  5 & 2.296 & 102 & 0.907 & 13 & 2.204 & 69 & 0.00122 & 107\\
 Journal of Philosophy of Education & 0.392 &  6 & 1.954 & 187 & 0.765 & 28 & 0.813 & 220 & 0.00062 & 185\\
 British Journal of Sociology of Education & 0.391 &  7 & 2.600 & 34 & 1.017 &  6 & 1.782 & 112 & 0.00246 & 46\\
 Oxford Review of Education & 0.380 &  8 & 2.551 & 44 & 0.969 & 10 & 1.421 & 159 & 0.00171 & 79\\
 Learning Media and Technology & 0.374 &  9 & 2.761 & 11 & 1.033 &  5 & 2.547 & 44 & 0.00133 & 98\\
 Comparative Education Review & 0.373 & 10 & 2.563 & 40 & 0.957 & 11 & 2.246 & 67 & 0.00126 & 103\\
 Review of Research in Education & 0.368 & 11 & 2.340 & 87 & 0.860 & 15 & 4.667 &  6 & 0.00137 & 94\\
 Curriculum Inquiry & 0.362 & 12 & 2.223 & 127 & 0.805 & 21 & 1.111 & 190 & 0.00066 & 179\\
 Asia Pacific Journal of Education & 0.358 & 13 & 1.948 & 188 & 0.697 & 39 & 0.733 & 228 & 0.00058 & 193\\
 International Journal of Computer-Supported Collaborative Learning & 0.354 & 14 & 2.866 &  5 & 1.015 &  7 & 4.028 &  9 & 0.00074 & 166\\
 Professional Development in Education & 0.348 & 15 & 2.796 &  9 & 0.974 &  9 & 1.531 & 138 & 0.00086 & 153\\
 Teachers and Teaching & 0.348 & 16 & 2.843 &  6 & 0.990 &  8 & 2.345 & 55 & 0.00214 & 53\\
 TESOL Quarterly & 0.341 & 17 & 3.103 &  1 & 1.060 &  3 & 2.071 & 84 & 0.00256 & 41\\
 Language Learning & 0.339 & 18 & 2.636 & 27 & 0.895 & 14 & 3.408 & 21 & 0.00379 & 17\\
 Journal of Education for Teaching & 0.336 & 19 & 2.018 & 180 & 0.678 & 44 & 1.483 & 149 & 0.00114 & 114\\
 Educational Researcher & 0.334 & 20 & 2.324 & 93 & 0.776 & 24 & 3.483 & 20 & 0.00660 &  7\\
 Compare-A Journal of Comparative and International Education & 0.331 & 21 & 2.837 &  7 & 0.940 & 12 & 1.607 & 127 & 0.00140 & 92\\
 Higher Education Research \& Development & 0.327 & 22 & 2.296 & 103 & 0.750 & 31 & 2.129 & 78 & 0.00297 & 29\\
 Teachers College Record & 0.325 & 23 & 2.211 & 131 & 0.718 & 35 & 0.970 & 205 & 0.00392 & 16\\
 Asia-Pacific Education Researcher & 0.325 & 24 & 1.868 & 204 & 0.606 & 59 & 0.744 & 227 & 0.00099 & 131\\
 Journal of Research on Educational Effectiveness & 0.317 & 25 & 2.531 & 48 & 0.802 & 22 & 3.375 & 22 & 0.00288 & 30\\
 Educational Research Review & 0.317 & 26 & 2.449 & 64 & 0.776 & 25 & 6.962 &  2 & 0.00299 & 28\\
 Cambridge Journal of Education & 0.312 & 27 & 2.639 & 26 & 0.823 & 18 & 1.421 & 159 & 0.00110 & 119\\
 Educational Philosophy and Theory & 0.311 & 28 & 2.359 & 81 & 0.733 & 33 & 0.773 & 222 & 0.00173 & 78\\
 Language Teaching & 0.309 & 29 & 2.632 & 28 & 0.814 & 20 & 3.714 & 11 & 0.00157 & 85\\
 Assessment \& Evaluation in Higher Education & 0.308 & 30 & 2.167 & 144 & 0.667 & 49 & 2.320 & 56 & 0.00213 & 54\\
 Journal of Literacy Research & 0.304 & 31 & 2.238 & 122 & 0.679 & 41 & 2.255 & 64 & 0.00089 & 150\\
 British Journal of Educational Technology & 0.301 & 32 & 2.715 & 16 & 0.817 & 19 & 2.951 & 31 & 0.00341 & 23\\
 International Journal of Science Education & 0.301 & 33 & 2.785 & 10 & 0.837 & 17 & 1.485 & 146 & 0.00416 & 15\\
 Pedagogische Studien & 0.292 & 34 & 1.402 & 230 & 0.409 & 123 & 0.245 & 261 & 0.00008 & 257\\
 Asia Pacific Education Review & 0.290 & 35 & 2.345 & 85 & 0.681 & 40 & 0.761 & 224 & 0.00064 & 180\\
 Computers \& Education & 0.288 & 36 & 2.967 &  2 & 0.854 & 16 & 5.296 &  4 & 0.01337 &  1\\
 Journal of Education Policy & 0.288 & 37 & 2.692 & 18 & 0.775 & 26 & 3.048 & 29 & 0.00267 & 35\\
 Mathematical Thinking and Learning & 0.288 & 38 & 2.651 & 23 & 0.762 & 29 & 1.074 & 195 & 0.00033 & 223\\
 Reading Research Quarterly & 0.287 & 39 & 2.338 & 89 & 0.672 & 48 & 3.543 & 17 & 0.00179 & 76\\
 Linguistics and Education & 0.287 & 40 & 2.626 & 30 & 0.753 & 30 & 1.289 & 176 & 0.00107 & 122\\
 American Educational Research Journal & 0.286 & 41 & 2.565 & 39 & 0.733 & 32 & 5.013 &  5 & 0.00571 &  9\\
 Education Finance and Policy & 0.286 & 42 & 2.373 & 76 & 0.678 & 45 & 2.395 & 51 & 0.00216 & 51\\
 English in Australia & 0.283 & 43 & 1.432 & 229 & 0.405 & 125 & 0.250 & 258 & 0.00005 & 260\\
 Advances in Health Sciences Education & 0.283 & 44 & 2.337 & 90 & 0.661 & 51 & 2.480 & 47 & 0.00421 & 14\\
 Journal of Higher Education Policy and Management & 0.282 & 45 & 2.489 & 54 & 0.703 & 37 & 0.939 & 208 & 0.00092 & 144\\
 AERA Open & 0.278 & 46 & 2.383 & 75 & 0.663 & 50 & 1.892 & 99 & 0.00273 & 33\\
 Language Policy & 0.278 & 47 & 2.512 & 50 & 0.698 & 38 & 1.383 & 165 & 0.00063 & 183\\
 Journal of Educational Change & 0.273 & 48 & 2.664 & 22 & 0.726 & 34 & 1.791 & 110 & 0.00077 & 165\\
 English in Education & 0.269 & 49 & 2.500 & 51 & 0.674 & 47 & 0.625 & 235 & 0.00015 & 248\\
 Journal of Psychologists and Counsellors in Schools & 0.267 & 50 & 2.248 & 117 & 0.601 & 60 & 0.676 & 231 & 0.00012 & 253\\
   \hline\hline
   \end{tabular}}
\end{table}

\begin{table}[ht]
\caption{Institutions with the greatest output of citing papers of \emph{JPCC}. CP represents the total number of citing papers affiliated a specific institution. $\rho$ is the percentage of citing papers affiliated a specific institution. PY denotes  the average publication year of CP.}
\label{Tb:Citingpaper:Institution}
\centering
   \begin{tabular}{llcccccccccccccccccccc}
   \hline\hline
  $R$ & Institution & CP &  $\rho$ &  PY  \\
   \hline
  1  & UCL & 10 & 2.78\% & 2017.6 \\
  2  & Univ Murcia &  6 & 1.67\% & 2018.7 \\
  3  & Univ Calif San Diego &  6 & 1.67\% & 2018.8 \\
  4  & Univ Twente &  5 & 1.39\% & 2018.4 \\
  5  & Univ Glasgow &  5 & 1.39\% & 2018.6 \\
  6  & Univ Antwerp &  5 & 1.39\% & 2017.6 \\
  7  & Univ Autonoma Barcelona &  5 & 1.39\% & 2018.4 \\
  8  & Boston Coll &  5 & 1.39\% & 2018.8 \\
  9  & Univ Oxford &  5 & 1.39\% & 2018.8 \\
 10  & Univ Amsterdam &  4 & 1.11\% & 2018.0 \\
 11  & Univ Manchester &  4 & 1.11\% & 2018.8 \\
 12  & Open Univ Cyprus &  4 & 1.11\% & 2019.2 \\
 13  & Univ Auckland &  4 & 1.11\% & 2019.8 \\
 14  & Michigan State Univ &  3 & 0.83\% & 2019.0 \\
 15  & Nanyang Technol Univ &  3 & 0.83\% & 2020.0 \\
 16  & Univ Texas Austin &  3 & 0.83\% & 2019.0 \\
 17  & Elon Univ &  3 & 0.83\% & 2019.3 \\
 18  & Leiden Univ &  3 & 0.83\% & 2018.7 \\
 19  & Univ Bucharest &  3 & 0.83\% & 2018.0 \\
 20  & Birkbeck Univ London &  3 & 0.83\% & 2017.0 \\
   \hline\hline
   \end{tabular}
\end{table}

\newpage


\end{document}